\documentclass[aps,prl,twocolumn,showpacs]{revtex4}
\usepackage{amsmath}
\usepackage{amsfonts}
\usepackage{amssymb}
\usepackage{epsfig}

\newcommand{\ra}{\rangle}
\newcommand{\la}{\langle}

\begin{document}

\title{Test of the Additivity Principle for Current Fluctuations \\
in a Model of Heat Conduction}

\author{Pablo I. Hurtado}
\author{Pedro L. Garrido}
\affiliation{
Instituto Carlos I de F\'{\i}sica Te\'orica y Computacional, Universidad de Granada, 
Granada 18071, Spain}

\date{\today}

\begin{abstract}
The additivity principle allows to compute the current distribution in many one-dimensional (1D) nonequilibrium systems. Using simulations, 
we confirm this conjecture in the 1D Kipnis-Marchioro-Presutti model of heat conduction for a wide current interval. The current distribution 
shows both Gaussian and non-Gaussian regimes, and obeys the Gallavotti-Cohen fluctuation theorem. 
We verify the existence of a well-defined temperature profile associated to a given current fluctuation. This profile is independent of the 
sign of the current, and this symmetry extends to higher-order profiles and spatial correlations. We also show that finite-time joint fluctuations
of the current and the profile are described by the additivity functional. These results suggest the additivity hypothesis as a general and 
powerful tool to compute current distributions in many nonequilibrium systems. 
\end{abstract}


\maketitle

Nonequilibrium systems typically exhibit currents of different observables (e.g., mass or energy)
which characterize their macroscopic behavior. Understanding how microscopic dynamics
determine the long-time averages of these currents and their fluctuations is one of the main objectives
of nonequilibrium statistical physics \cite{BD,Bertini,Derrida,Livi,we,GC,LS}. This problem has proven to be a 
challenging task, and up to now only few exactly-solvable cases are understood \cite{BD,Bertini,Derrida}.
An important step in this direction has been the development of the Gallavotti-Cohen fluctuation theorem \cite{GC,LS}, 
which relates the probability of forward and backward currents reflecting the time-reversal symmetry of microscopic dynamics. 
However, we still lack a general approach based on few simple principles.
Recently, Bertini and coworkers \cite{Bertini} have introduced a Hydrodynamic Fluctuation Theory (HFT) 
to study large dynamic fluctuations in nonequilibrium steady states. This is a very general approach which leads to a hard
optimization problem whose solution remains challenging in most cases.
Simultaneously, Bodineau and Derrida \cite{BD} have conjectured an additivity principle for current fluctuations in 1D
which can be readily applied to obtain quantitative predictions and, together with HFT,
seems to open the door to a general theory for nonequilibrium systems.

The additivity principle (also referred here as BD theory) enables one to calculate the fluctuations of the current in 1D diffusive 
systems in contact with two boundary thermal baths at different temperatures, $T_L\neq T_R$.
It is a very general conjecture of broad applicability, expected to hold for 1D systems of classical interacting particles,
both deterministic or stochastic, independently of the details of the interactions between the particles or the coupling to the thermal reservoirs. 
The only requirement is that the system at hand must be diffusive, i.e. Fourier's law must hold. If this is the case, the additivity principle predicts 
the full current distribution in terms of its first two cumulants.
Let $\text{P}_N(q,T_L,T_R,t)$ be the probability of observing a time-integrated current $Q_t=q t$ during a long 
time $t$ in a system of size $N$. This probability obeys a large deviation principle \cite{LD}, $\text{P}_N(q,T_L,T_R,t)\sim \exp[t {\cal F}_N(q,T_L,T_R)]$, 
where ${\cal F}_N(q,T_L,T_R)$ is the current large-deviation function (LDF), 
meaning that current fluctuations away from the average are exponentially unlikely in time.
The additivity principle relates this probability with the 
probabilities of sustaining the same current in subsystems of lengths $N-n$ and $n$, 
and can be written as
${\cal F}_{N}(q,T_L,T_R) = \max_T \left[ {\cal F}_{N-n}(q,T_L,T) + {\cal F}_{n}(q,T,T_R) \right]$ for the LDF \cite{BD}. 
In the continuum limit one gets \cite{BD}
\begin{equation}
{\cal G}(q)=- \min_{T_q(x)}\left\{ \int_0^1 \frac{\left[q + \kappa[T_q(x)] T'_q(x)\right]^2}
{2\sigma[T_q(x)]} \text{d}x \right\} \, , 
\label{ldf1}
\end{equation}
with ${\cal G}(q)=N {\cal F}_N(\frac{q}{N})$ -- we drop the dependence on the baths for convenience --, $x\in [0,1]$, and where
$\kappa(T)$ is the thermal conductivity appearing in Fourier's law, $\langle Q_t\rangle/t=-\kappa(T)\, \nabla T$, 
and $\sigma(T)$ measures current fluctuations in equilibrium ($T_L=T_R$), $\langle Q_t^2\rangle/t=\sigma(T)/N$.
The optimal profile $T_q(x)$ derived from (\ref{ldf1}) obeys
\begin{equation}
\kappa^2[T_q(x)] \left(\frac{\text{d} T_q(x)}{\text{d}x}\right)^2 = q^2\left\{1+2K \sigma[T_q(x)] \right\}  \, ,
\label{optprof}
\end{equation}
where $K(q^2)$ is a constant which fixes the correct boundary conditions, $T_q(0)=T_L$ and $T_q(1)=T_R$. 
Eqs. (\ref{ldf1}) and (\ref{optprof}) completely determine the current distribution, 
which is in general non-Gaussian and obeys the Gallavotti-Cohen symmetry, i.e. ${\cal G}(-q) = {\cal G}(q) -E\, q$ with $E$ some constant 
defined by $\kappa(T)$ and $\sigma(T)$ \cite{BD}. 

The additivity principle is better understood within the context of HFT \cite{Bertini}, which
provides a variational principle for the most probable (possibly time-dependent)
profile responsible of a given current fluctuation, leading usually to unmanageable equations.
The additivity principle, which on the other hand yields explicit predictions,
is equivalent within HFT to the hypothesis that
the optimal profile is time-independent, an approximation which in some special cases breaks down for for extreme 
current fluctuations \cite{Bertini}. Even so, the additivity principle correctly predicts the current LDF 
in a very large current interval, making it very appealing.

The goal of this paper is to test the additivity principle in a particular system: the 1D Kipnis-Marchioro-Presutti (KMP) 
model of heat conduction \cite{kmp}. The model is defined on a 1D open lattice 
with $N$ sites. A configuration is given by $\{e_i, i=1\ldots N \}$, where $e_i\in \mathbb{R}_+$ is the energy of  site $i$. 
Dynamics is stochastic, proceeding through random energy exchanges between randomly-chosen nearest neighbors.
In addition, boundary sites ($i=1,N$) may also exchange energy with boundary heat baths whose energy is randomly drawn 
at each step from a Gibbs distribution at the corresponding temperature ($T_L$ for $i=1$ and $T_R$ for $i=N$). For $T_L \neq T_R$ KMP
proved that the system has a steady state characterized by an average current $\la q \ra$ and a linear energy profile 
$T_{\text{st}}(x)=T_L + x(T_R - T_L)$ in the $N\to \infty$ hydrodynamic scaling limit, such that Fourier's law holds with
$\kappa(T)=\frac{1}{2}$. Moreover, one also gets $\sigma(T)=T^2$. This model plays a fundamental role in nonequilibrium statistical 
physics as a benchmark to test new theoretical advances, and represents a large class of quasi-1D diffusive systems of technological 
and theoretical interest. Furthermore, the KMP model is an optimal candidate to test
the additivity principle because: (i) One can obtain explicit predictions for its current LDF,
and (ii) its simple dynamical rules allow a detailed numerical study of current fluctuations.

\begin{figure}[t]
\centerline{
\psfig{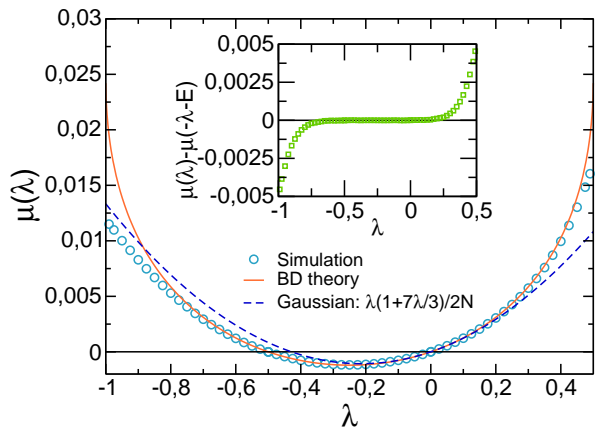}}
\caption{(Color online) Main: $\mu(\lambda)$ for the 1D KMP model. Fluctuations are Gaussian
for $\lambda\approx 0$, but non-Gaussian in the tails. Inset: Test of the Gallavotti-Cohen fluctuation relation.}
\label{ldf}
\end{figure}

Analytical expressions for ${\cal G}(q)$ and $T_q(x)$ in the KMP model, though lengthy, are easily derived from 
the additivity principle, see eqs. (\ref{ldf1}) and (\ref{optprof}), and will be published elsewhere \cite{Pablo}. We find that 
$\text{P}_N(q,T_L,T_R,t)$ is Gaussian around $\la q \ra$ with variance $\sigma(T)$, while 
non-Gaussian tails develop far from $\la q \ra$. Exploring by standard simulations these tails to check BD theory is unapproachable,
since LDFs involve by definition exponentially-unlikely rare events.
Recently  Giardin\`a, Kurchan and Peliti \cite{sim} have introduced an efficient method to measure LDFs in many particle 
systems, based on a modification of the dynamics so that the rare events responsible of the large deviation are 
no longer rare \cite{sim2}. This method yields the Legendre transform of the current LDF, 
$\mu(\lambda)\equiv N^{-1}\max_q[{\cal G}(q) + \lambda q ]$. The function $\mu(\lambda)$ can be viewed as the conjugate \emph{potential}  
to ${\cal G}(q)$, with $\lambda$ the parameter conjugate to the current $q$, a relation equivalent to the free energy being the Legendre transform 
of the internal energy in thermodynamics, with the temperature as conjugate parameter to the entropy.

We applied the method of Giardin\`a et al to measure $\mu(\lambda)$ for the 1D KMP model with $N=50$, $T_L=2$ and $T_R=1$, see Fig. \ref{ldf}. 
The agreement with BD theory is excellent for a wide $\lambda$-interval, say $-0.8<\lambda<0.45$, which corresponds to a very large range 
of current fluctuations. Moreover, the deviations observed for extreme current fluctuations are due to known limitations of the algorithm \cite{Pablo}, so 
no violations of additivity are observed.
In fact, we can use the Gallavotti-Cohen symmetry, 
$\mu(\lambda)=\mu(-\lambda-E)$ with $E= (T_R^{-1}-T_L^{-1})$, to bound the range of validity of the algorithm. The inset to Fig. \ref{ldf}
shows that this symmetry holds in the large current interval for which the additivity principle predictions agree with measurements,
thus confirming its validity in this range.
However, we cannot discard the possibility of an additivity breakdown for extreme current fluctuations due to the onset of time-dependent 
profiles \cite{Bertini}, although we stress that such scenario is not observed here. 

\begin{figure}[t]
\centerline{
\psfig{file=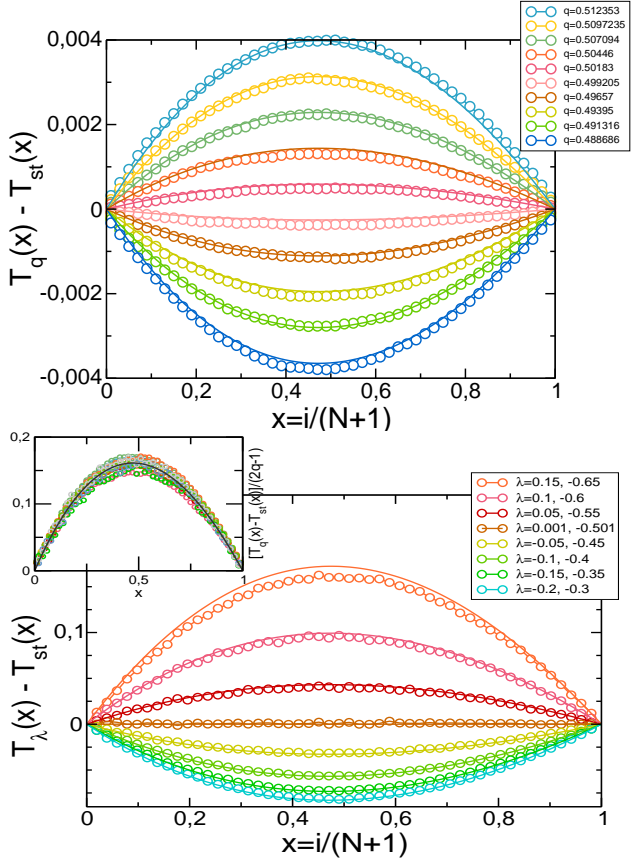,width=8.3cm}}
\caption{(Color online) Excess profiles for different currents measured in standard simulations (top) 
and with the method of \cite{sim} (bottom). 
Much larger current fluctuations can be explored in the latter case. 
The inset shows a scaling plot of excess profiles for small current deviations, $q\in [0.44,0.62]$. 
In all cases lines correspond to BD theory.}
\label{prof}
\end{figure}

The additivity principle leads to the minimization of a functional of the temperature profile, $T_q(x)$, see  eqs. (\ref{ldf1}) and (\ref{optprof}).
A relevant question is whether this optimal profile is actually observable. We naturally define $T_q(x)$ as the average \emph{energy} 
profile adopted by the system during a large deviation event of (long) duration $t$ and time-integrated current $q$, measured at an 
\emph{intermediate time} $1\ll \tau \ll t$. Top panel in Fig. \ref{prof} shows $T_q(x)$ measured in standard simulations for small
current fluctuations, and the agreement with BD predictions is again very good. This confirms the idea that the system modifies  
its temperature profile to facilitate the deviation of the current. 

To obtain optimal profiles for larger current fluctuations we may use
the method of Giardin\`a et al \cite{sim}. This method naturally yields $T_{\lambda}^{\text{end}}(x)$, the average energy profile \emph{at the end} of the 
large deviation event ($\tau=t$), which can be connected to the correct observable $T_{\lambda}(x)$  
by noticing that KMP dynamics obeys the local detailed balance condition \cite{LS}, which guarantees the time 
reversibility of microscopic dynamics. This condition implies a symmetry between the forward dynamics for a current fluctuation and the 
time-reversed dynamics for the negative fluctuation \cite{Rakos} that can be used to derive the following relation between midtime and 
endtime statistics \cite{Pablo}
\begin{equation}
P_{\lambda}(C) = A \times (p_C^{\text{eff}})^{-1} P_{\lambda}^{\text{end}}(C) P_{-\lambda-E}^{\text{end}}(C) \, .
\label{probmid}
\end{equation}
Here $P_{\lambda}^{\text{end}}(C)$ [resp. $P_{\lambda}(C)$] is the probability of configuration $C$ at the end (resp. at 
intermediate times) of a large deviation event with current-conjugate parameter $\lambda$, and 
$p_{C}^{\text{eff}}\simeq \exp [-\sum_{i=1}^N \beta_i e_i ]$ is an effective weight for configuration 
$C=\{e_i, i=1\ldots N\}$, with $\beta_i=T_L^{-1} + E \frac{i-1}{N-1}$, while $A$ is a normalization constant.
Eq. (\ref{probmid}) implies that configurations with a significant contribution to the average profile at intermediate times are 
those with an important probabilistic weight at the end of both the large deviation event and its time-reversed process. 
Using the relation (\ref{probmid}) and a local equilibrium (LE) approximation \cite{Pablo}, we obtained $T_{\lambda}(x)$ for a large interval 
of current fluctuations using the method of Giardin\`a et al \cite{sim},
see Fig. \ref{prof} (bottom). The excellent agreement with BD profiles confirms 
the additivity principle as a powerful conjecture to compute both the current LDF and the associated optimal profiles. Moreover, this agreement shows
that corrections to LE are weak in the KMP model, though we show below that these small corrections are present and can be measured.
\begin{figure}[t]
\centerline{
\psfig{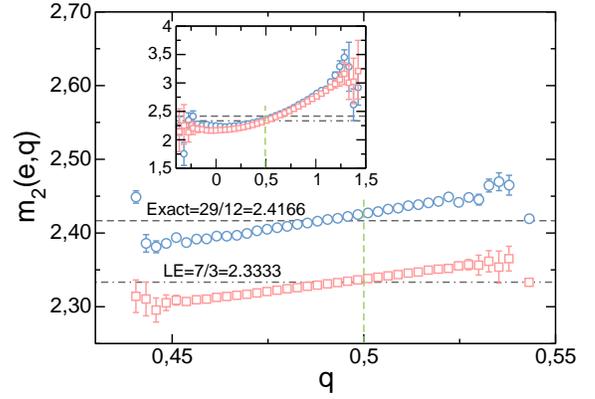}}
\caption{(Color online) Main: Fluctuations of total energy vs $q$ measured in standard simulations for $t=10^6$ ($\bigcirc$) and LE results ($\Box$).
Inset: Similar results for $t=4000$. 
Notice the non-trivial structure.
}
\label{ener}
\end{figure}

An important consequence of eq. (\ref{probmid}) is that $P_{\lambda}(C)=P_{-\lambda-E}(C)$, or equivalently $P_q(C)=P_{-q}(C)$, so midtime statistics
does not depend on the sign of the current. This implies in particular that $T_q(x)=T_{-q}(x)$, but also that all higher-order profiles and spatial 
correlations are independent of the current sign. 

As another test, BD theory predicts for $q\approx \la q\ra=\frac{1}{2}$ the limiting behavior
\begin{equation}
\frac{T_q(x)-T_{\text{st}}(x)}{2q-1} = \frac{1}{7} x (1-x) (5-x) + {\cal O}(2q-1) \, . \nonumber
\label{limiting}
\end{equation}
The inset to Fig. \ref{prof} confirms this scaling for $T_q(x)$ and many different values of $q$ around its average.

We can now go beyond the additivity principle by studying fluctuations of the system total energy, for which current theoretical approaches cannot 
offer any prediction. An exact result by Bertini, Gabrielli and Lebowitz \cite{BGL} predicts that $m_2(e) = m_2^{LE}(e) + \frac{1}{12}(T_L-T_R)^2$, 
where $m_2(e)$ is the variance of the total energy, $m_2^{LE}$ is the variance assuming a local equilibrium (LE) product measure, 
and the last term reflects the correction due to the long-range correlations in the nonequilibrium stationary state \cite{BGL}. In our case, 
$m_2^{LE}=(T_L^2 + T_LT_R + T_R^2)/3=7/3 \approx 2.3333$, while $m_2=29/12 \approx 2.4166$. 
Fig. \ref{ener} plots $m_2(e,q)=N[\la e^2\ra(q) - \la e\ra^2(q)]$ measured in standard simulations, showing a non-trivial, interesting structure for $m_2(e,q)$ 
which both BD theory and HFT cannot explain. One might obtain a theoretical prediction for $m_2(e,q)$ by supplementing
the additivity principle with a LE hypothesis,  i.e. $P_q(C)\approx \Pi_{i=1}^N \exp[-e_i/T_q(\frac{i}{N+1})]$, which results in
$m_2^{LE}(e,q) = \int_0^1 \text{d} x \, T_q(x)^2$. However, Fig. \ref{ener} shows that $m_2^{LE}(e,\la q \ra)\approx 2.33$ as corresponds to a LE picture, and 
in contrast to the exact result $m_2(e,\la q \ra)\approx 2.4166$. This proves that, even though LE is a sound numerical hypothesis to obtain $T_{\lambda}(x)$ 
from endtime statistics using the method of Giardin\`a et al \cite{sim,Pablo}, see Fig. \ref{prof} (bottom), corrections to LE become apparent at the fluctuating level. 

\begin{figure}[t]
\centerline{
\psfig{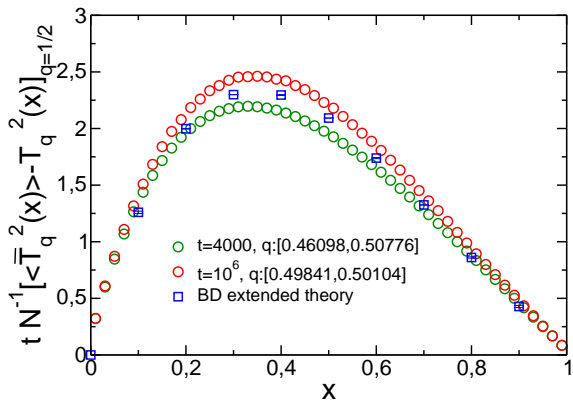}}
\caption{(Color online) Main: Finite-time profile fluctuations for $N=50$ and extended BD prediction.
}
\label{joint}
\end{figure}

For long but finite times, the profile associated to a given current fluctuation is subject to fluctuations itself. 
These joint fluctuations of the current and the profile are again not described by the additivity principle, but we may study them
by extending the additivity conjecture. In this way, we now assume that the probability to find a time-integrated current
$q/N$ \emph{and} a temperature profile $\bar{T}_q(x)$ after averaging for a long but finite time $t$ can be written as
${\bar P}_N[\frac{q}{N},\bar{T}_q(x);t] \sim \exp\{\frac{t}{N} \bar{\cal G}[q,\bar{T}_q(x)]\}$, where $\bar{\cal G}[q,\bar{T}_q(x)]$ 
is the functional of eq. (\ref{ldf1}) but not subject to the minimization procedure (with the notation change $T_q \to \bar{T}_q$). In this scheme the profile 
obeying eq. (\ref{optprof}), i.e. the one which minimizes the functional $\bar{\cal G}$, is the \emph{classical} profile
$T_{q}(x)$. We can make a perturbation of $\bar{T}_q(x)$ around its classical value,
$\bar{T}_q(x)=T_{q}(x)+\eta_q(x)$, and for long $t$ the joint probability of $q$ and $\eta_q(x)$ obeys
\begin{equation}
\frac{{\bar P}_N[q,\eta_q(x);t]}{P_N(q,t)} \simeq \exp \left[ -\frac{1}{2} \int_0^1 \text{d}x \text{d}y A_q(x,y) \eta_q(x) \eta_q(y) \right] \, , \nonumber
\label{adext}
\end{equation}
where the $xy$-symmetric kernel is 
$A_q(x,y)=t (4 N T_{q}^2)^{-1} \left[ 2 T_{q}^{-1} T'_{q} \partial_x - \partial_x^2 -8q^2 K(q^2) \right]\delta(x-y)$.
In order to check this approach, we studied the observable $\la \bar{T}_q^2(x)\ra - T_{q}^2(x)=\la\eta_q^2(x)\ra=A_q^{-1}(x,x)$.
The function $A_q^{-1}(x,x)$ can be written in terms of the eigenvectors and eigenvalues of kernel $A_q(x,y)$ \cite{Pablo}.
For the particular case $q=\la q\ra=\frac{1}{2}$ we were able to solve analytically 
this problem in terms of Bessel functions,
leading to an accurate numerical evaluation of $A_q^{-1}(x,x)$ \cite{Pablo}. We compare these results in Fig. \ref{joint} with standard
simulations for both $t=4\times 10^3$ and $t=10^6$. Good agreement is found in both cases, pointing out that BD functional $\bar{\cal G}[q,\bar{T}_q(x)]$
contains the essential information on the joint fluctuations of the current and profile. 

In summary, we have confirmed the additivity principle in the 1D KMP model of heat conduction for a large current interval, extending its validity 
to joint current-profile fluctuations. These results strongly support the additivity principle as a general 
and powerful tool to compute current distributions in many 1D nonequilibrium systems, opening the door to a general approach 
based on few simple principles. Our confirmation does not discard however the possible 
breakdown of additivity for extreme current fluctuations due to the onset of time-dependent profiles, although we stress that this scenario is not 
observed here and would affect only the far tails of the current distribution. In this respect it would be interesting to study the KMP model on a 
ring, for which a dynamic phase transition to time-dependent profiles is known to exist \cite{Bertini}. Also interesting is the possible extension of 
the additivity principle to low-dimensional systems with anomalous, non-diffusive transport properties \cite{we}, or to systems with several conserved 
fields or in higher dimensions.

We thank  B. Derrida, J.L. Lebowitz,V. Lecomte and J. Tailleur for illuminating discussions.
Financial support from University of Granada and AFOSR Grant AF-FA-9550-04-4-22910 is also acknowledged.

\end{document}